\documentclass[twocolumn]{aastex701}
\usepackage{amsmath}

\newcommand{\msun}{M_{\odot}}
\newcommand{\rsun}{R_{\odot}}
\newcommand{\tdyn}{t_{\rm dyn}}
\newcommand{\Ebind}{|E_{\rm bind}|}
\newcommand{\Enuc}{E_{\rm nuc}}
\newcommand{\Eenc}{E_{\rm enc}}

\newcommand{\Edf}{E_{\rm DF}}
\newcommand{\bbeta}{\bar{\beta}}
\newcommand{\athenak}{\texttt{AthenaK}}
\newcommand{\mbh}{M_{\bullet}}
\newcommand{\mstar}{M_{\star}}
\newcommand{\rstar}{R_{\star}}

\shorttitle{Nuclear burning in BH--star collisions}
\shortauthors{Jiang \& Lin}

\begin{document}

\title{Stellar Disruption and Black Hole Retention in Massive Collisions: The Role of Thermonuclear Feedback}

\author[orcid=0000-0003-0292-2773]{Hong-Xuan Jiang}
\email{hongxuan\_jiang@sjtu.edu.cn}
\affiliation{The Hong Kong Institute for Astronomy and Astrophysics, The University of Hong Kong, Pokfulam Road, Hong Kong, China}
\affiliation{Tsung-Dao Lee Institute, Shanghai Jiao Tong University, Shengrong Road 520, Shanghai, 201210, China}

\author[orcid=0000-0001-5466-4628]{Douglas N. C. Lin}
\email{dnlin@ucsc.edu, lin@ucolick.org}
\affiliation{Department of Astronomy and Astrophysics, University of California, Santa Cruz, CA, 95064, USA}
\affiliation{Department of Astronomy, Westlake University, Hangzhou, 310024, Zhejiang, China}
\affiliation{Institute for Advanced Studies, Tsinghua University, Beijing, 100086, China}

\begin{abstract}
A stellar-mass black hole (BH) that collides with a massive star can be captured by drag and settle at its center as a black hole star. Hydrodynamic simulations that neglect nuclear reactions have found that the star survives and becomes a black hole star (BH*) for BH-to-star mass ratios $q\lesssim0.2$. We perform hydrodynamic simulations of head-on collisions with the \athenak\ code, including CNO burning coupled to the gas, for $q=0.1$ and $0.2$ BHs, stellar masses of $100$ and $500\,\msun$, solar,
metal-deficient, and Helium-rich compositions. In all models (with $q\le0.2$), the energy dissipated by drag is comparable 
to or less than the binding energy that the captured BH adds to the gas so that it alone does not disrupt the star.
But, under compression around the captured BH, gas is heated 
to $\gtrsim2\times10^8$\,K and undergoes thermonuclear 
burning at the $\beta$-limited hot-CNO rate.  Within about one dynamical time after the BH settles at the center, the nuclear energy released during the encounter becomes a significant fraction of the star's binding energy $\tfrac32\bbeta\int P\,dV$. The star loses most of its mass once the cumulative release 
reaches this binding energy. Below
this threshold, the release rate declines. Merger 
between a solar-composition star and a BH with a mass ratio
$q=0.2$ leads to the disruption of the star after
releasing most of its total mass. But a metal-poor star,
with nearly the same binding energy and total yield, survives 
with minor release after merger with an identical BH. 
After coalescence with a less massive ($q=0.1$) BH, stars with both solar and metal-poor composition
survive as black hole stars, burning at a rate 
$\gtrsim 10^{44}\,{\rm erg\,s^{-1}}$, far from thermal equilibrium. Very massive and helium-rich stars, which are weakly bound, are the most easily disrupted.
\end{abstract}

\keywords{Stellar mass black holes (1611) --- Massive stars (732) --- Stellar dynamics (1596) --- Nuclear astrophysics (1129) --- Hydrodynamical simulations (767)}

\section{Introduction}
\label{sec:intro}

Dense stellar systems bring stellar-mass black holes (BHs) into direct contact with massive stars. In young and nuclear star clusters the most massive stars, built by runaway mergers, are the most frequent collision targets \citep{PortegiesZwart2002,Kremer2019,Rose2022,Kiroglu2025,Rastello2026,Rantala2026}. In the disks of active galactic nuclei (AGN), gas drag traps stars and their BH remnants in a thin layer while stars grow to hundreds of solar masses \citep{2026ApJ...997..206X,Cantiello2021,Jermyn2022,AliDib2023,Chen2024}. Collisions 
between members of these two populations
that neither disrupt the star nor let the BH escape, 
leave BHs embedded in the stellar envelope. Such merger products are the BH analogues of Thorne--\.Zytkow objects \citep{Thorne1977,Hirai2022} and of quasi-stars \citep{Begelman2008,Ball2011,Coughlin2024}. They have been invoked as hosts of hierarchical BH mergers \citep{Jermyn2022,Hu2026,Ma2026,Cantiello2026} and, at higher BH masses, as models for the little red dots seen by JWST \citep{Matthee2024,Naidu2025,Begelman2026,Inayoshi2026}.

These encounters are direct collisions, not tidal disruptions. A BH lighter than the star has its tidal radius, $\rstar(\mbh/\mstar)^{1/3}$, inside the star. It therefore penetrates the envelope and is decelerated by drag. If the star 
disintegrates, it is destroyed by the energy the BH deposits rather than by tides \citep{Perets2016,Kremer2022}.

\citet{Shi2026} find with simulations of a $100\,\msun$ polytrope that drag retains the BH for speeds at infinity below about $2\,[G(\mstar+\mbh)/\rstar]^{1/2}$ and that the drag energy exceeds the binding energy of the star once $\mbh\gtrsim0.2\,\mstar$. \citet{Ma2026} followed the encounters between a BH and 
a star with a mass ratio $1{:}34$ and found the star remains
intact, though it is significantly heated.
These calculations treat the gas as inert. So did the star--star collision models of \citet{Lai1993}, in which the shocked gas is compressed by less than a factor of two and stays hot too briefly for nuclear energy to affect the flow. The center of a $100\,\msun$ main sequence star has 
$T_{\rm c}\simeq3.7\times10^7$\,K and burns hydrogen through the CNO cycle, whose rate scales as $T^{15}$ there. 
During a BH post-capture inspiral, gas drag on its motion
compresses the gas pressure around it by nearly 
a hundred fold and increases the local temperature 
to $\gtrsim2\times10^8$\,K.  The nuclear reaction cycle
saturates at its $\beta$-limited hot-CNO rate and burns only $10^{-3}$ of the star's hydrogen, $0.07\,\msun$. 
It has the potential to release sufficient energy to 
unbind a $100\,\msun$ main sequence star. 

Here we analyze how this transitory enhancement of 
nuclear-energy release may affect the structure of the star.
In \S\ref{sec:methods}, we describe the computational 
method used in this analysis.  We present the results of
our simulations in \S\ref{sec:results} and discuss their 
implications in \S\ref{sec:discussion}.

\section{Numerical method and stellar models}
\label{sec:methods}

In this work, the simulations are performed with the GPU-accelerated code \athenak\ \citep{Stone2026,Stone2020}, using the tidal disruption framework of \citet{Jiang2026}. This framework includes multigrid self-gravity and a tabulated equation of state (EOS) for H/He chemistry and Local Thermodynamic Equilibrium radiation pressure \citep{Chabrier2021,Tomida2013,Timmes2000}. MeshBlock-level timestep subcycling is also used to speed up the simulations. Technical details are given in \citet{Jiang2026}.

\subsection{Stellar models}

Each progenitor is constructed in a hydrostatic equilibrium using the same tabulated EOS and self-gravity solver as in \citet{Jiang2026}. We adopt a two-zone stellar model
\citep{Kohler2015} including an adiabatic convective 
core which contains the innermost $80\%$ of the mass 
($95\%$ for the $500\,\msun$ model) and a
surrounding $n \simeq 3$ radiative envelope. 
The central concentration, $\rho_{\rm c}/\langle\rho\rangle=28$--$35$, is similar to that of the $n=2.75$ polytrope of \citet{Shi2026}. For our fiducial $100\,\msun$ model, the central density and temperature are $\rho_{\rm c}=1.08\,{\rm g\,cm^{-3}}$ and $T_{\rm c}=3.7\times10^7$\,K, respectively. The radii follow from Eddington's standard model \citep{Eddington1926}. In this model, the ratio of gas pressure to total pressure, $\beta$, is fixed by the mass and mean molecular weight. The stellar radius 
$R_\star$ is then set by requiring the nuclear luminosity to equal $4\pi cG\mstar(1-\beta)/\kappa$. For the solar-metallicity $100\,\msun$ star this prescription gives $R_\star \sim 16.5\,\rsun$, close to the $16\,\rsun$ of \citet{Shi2026}. We adopt $16\,\rsun$ for the $100\,\msun$ zero-age main-sequence star and $42\,\rsun$ for the $500\,\msun$ star. Time is measured in units of the dynamical time $\tdyn=(\rstar^3/G\mstar)^{1/2}$, which is $2.8$\,hr for the fiducial star.

For a star in a hydrostatic equilibrium, its gravitational 
energy is $W=-3\int P\,dV$, its internal energy is $U=\tfrac32\int P_{\rm gas}dV+3\int P_{\rm rad}dV$, 
and its binding energy is 
\begin{equation}
\Ebind = -(W+U) = \tfrac32\int P_{\rm gas}\,dV=\tfrac32\,\bbeta\int P\,dV ,
\label{eq:virial}
\end{equation}
where $\bbeta$ is the pressure-weighted gas fraction. This is the total energy $\tfrac12\beta W$ of the Eddington standard model \citep{Lai1993}, generalized to a star whose $\beta$ varies with radius. Radiation pressure supports a massive star against gravitational contraction but it does not contribute to its gravitational binding energy. Consequently, helium-rich stars (with higher $\mu$) and very massive stars (with lower $\beta$) are only weakly bound. Four compositions are included in this work (Table~\ref{tab:runs}). The solar mixture ($X=0.70$, $Y=0.28$, $Z=0.02$) is adopted as the fiducial composition. A metal-deficient SMC mixture \citep[$Z=0.0021$,][]{Brott2011} varies the CNO catalyst, while helium-enriched mixtures 
(to represent advanced stages of main sequence evolution)
with $Y=0.43$ and $0.78$ for fixed $Z$ values \citep{Yoon2006,Szecsi2015} vary the amount of nuclear
fuel as well as $\mu$ for a fixed catalyst.

\subsection{Nuclear burning model}
\label{sec:burning}

Hydrogen and helium mass fractions are carried as passive scalars and burned with a two-channel network. The CNO rate follows the standard cold-cycle expression \citep{Caughlan1988,Kippenhahn2012},
\begin{equation}
\epsilon_{\rm CNO}=8.24\times10^{25}\,\rho\,X_{\rm H}X_{\rm CNO}\,T_9^{-2/3}\,e^{-15.231\,T_9^{-1/3}},
\end{equation}
in ${\rm erg\,g^{-1}\,s^{-1}}$, where $\rho$ is in ${\rm g\,cm^{-3}}$, $X_{\rm H}$ and $X_{\rm CNO}$ are the mass fractions of hydrogen and CNO nuclei, and $T_9=T/10^9\,{\rm K}$. It is capped at the $\beta$-decay-limited hot-CNO value of $5.8\times10^{15}X_{\rm CNO}\,{\rm erg\,g^{-1}\,s^{-1}}$ \citep{Wallace1981}.
The triple-$\alpha$ rate is
\begin{equation}
\epsilon_{3\alpha}=5.09\times10^{11}\,\rho^2Y^3T_8^{-3}\exp(-44.027/T_8)
\end{equation}
where $T_8 = T/10^8$\,K.
The released energy is fed back into the gas at every step and accumulated in advected scalars. We set $X_{\rm CNO}=Z$, which overestimates the catalyst abundance of a scaled-solar mixture by a factor of about $1.4$. The CNO rate is linearly proportional to $X_{\rm CNO}$, but the resulting yield from the enhanced nuclear
reaction is not (\S\ref{sec:clamp}).
In the controlled no-feedback models,
the nuclear network is still computed to take into account 
the accelerated depletion of the hydrogen fuel, albeit 
no enhanced heat is deposited in the gas.
Above $1.5\times10^8$\,K, the CNO rate is limited by 
the $\beta$-decay lifetimes of the catalyst, so nuclear heating is far too slow to detonate the gas. The triple-$\alpha$ and pp channels are negligible under all conditions reached in our 
simulations.

\begin{figure*}[!t]
\centering
\includegraphics[width=\textwidth]{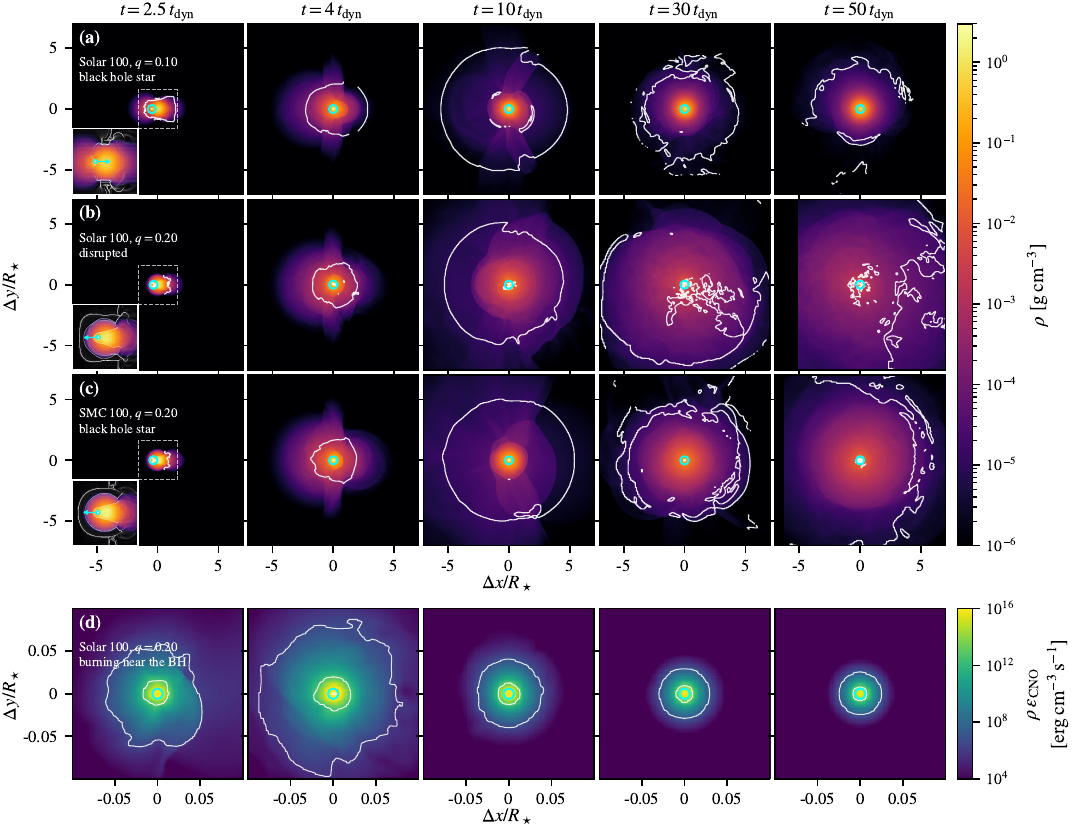}
\caption{Density in the orbital plane for the fiducial star hit by a $10\,\msun$ BH (a) and a $20\,\msun$ BH (b), and for the SMC star hit by a $20\,\msun$ BH (c), all with nuclear burning. The circle marks the BH, and the white contour encloses gas bound by its own gravity (Appendix~\ref{app:convergence}). Panels are centered on the densest gas, and the axes give offsets in units of $\rstar$. Insets zoom into the dashed box ($\pm1.6\,\rstar$); white lines mark shocks ($|\nabla\ln P|$) and the arrow shows the BH motion. Row~(d): nuclear heating rate per volume near the BH in run~(b), with density contours at $1$, $10$, and $100\,{\rm g\,cm^{-3}}$ and the softening sphere at the center.}
\label{fig:process}
\end{figure*}

\begin{figure*}[!t]
\centering
\includegraphics[width=.75\textwidth]{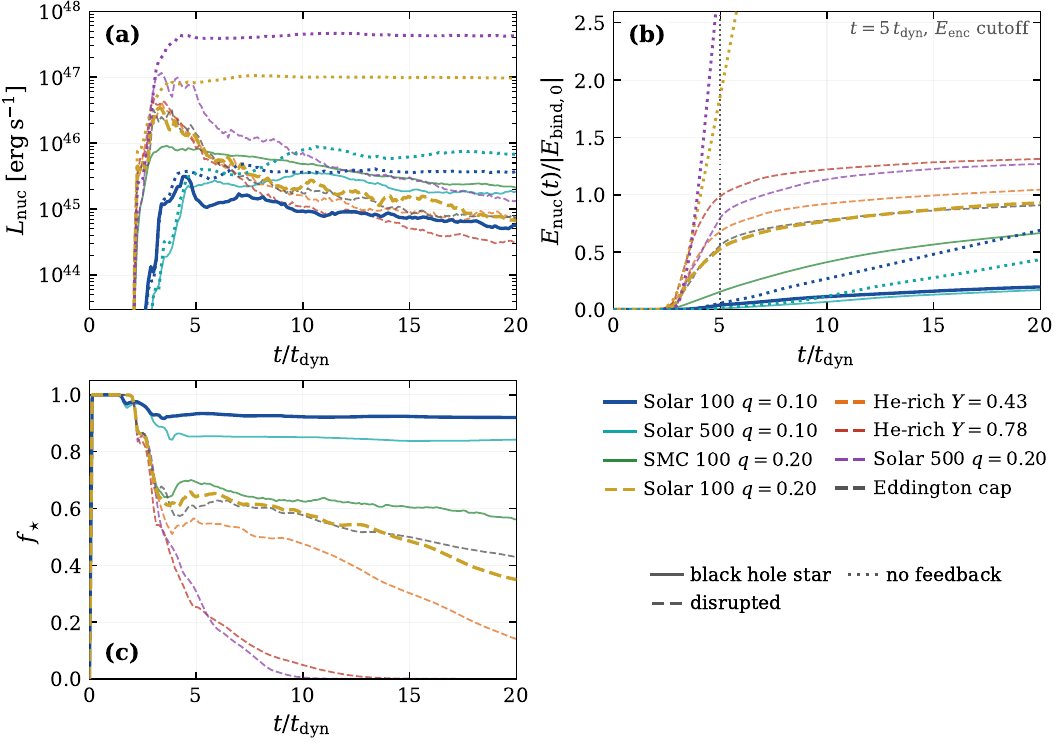}
\caption{Histories over the first $20\,\tdyn$: (a) nuclear heating rate outside the softening sphere, (b) cumulative release in units of $\Ebind$, with the $5\,\tdyn$ cutoff that defines $\Eenc$, and (c) $f_\star$. Solid lines end as black hole stars, dashed lines are disrupted, and dotted lines are the no-feedback controls (totals in Table~\ref{tab:full}). ``Eddington cap'' is the test run of Appendix~\ref{app:convergence}.}
\label{fig:histories}
\end{figure*}

\subsection{Collision setup and resolution}
\label{sec:setup}

The BH's potential is modeled as that of a Plummer-softened point 
mass with softening length $r_{\rm soft}=\rstar/160$.  During its
inspiral (on the dynamical timescale of a few hours), the 
mass it accretes  from the stellar envelope is negligible even 
at a rate far above BH's accretion rate capped by the 
Eddington limit.  Moreover, the frictional energy 
dissipation rate due to the hydrodynamic drag also greatly 
exceeds the Eddington limit.  Therefore, we neglect the effects
of BH's accretion.  

The BH is released on a radial orbit from 5 tidal radii away from the stellar center with the free-fall speed from rest at infinity. The simulation domain is $25\,\rstar$ wide and uses a $32^3$ root grid with static refinement in the central region and 9 levels of adaptive refinement on top. The star is resolved down to the length scale of $\rstar/82$, and the region around the BH is refined to the length scale of $\rstar/655$ ($0.024\,\rsun$ for the star in the fiducial model), so that $r_{\rm soft}$ is covered by 4.1 cells.
Within $r_{\rm soft}$, the softened potential causes gas to pile into an artificial cusp whose temperature and density are not reliably determined. We therefore switch off nuclear burning there, so the nuclear energies we quote are lower limits. Neglecting accretion onto the BH assumes that it removes 
negligible amount of the hot gas during the encounter.
At the peak of the enhanced nuclear burning episode, 
the burning region is centered around $1.3$--$1.9\, r_{\rm soft}$ from the BH, outside the softened sphere.
Appendix~\ref{app:convergence} lists some 
tests of these choices.

We track three energies, all in units of $\Ebind$, the binding energy of the initial model, which we hold fixed throughout.
The drag energy $\Edf$ is the maximum value of 
orbital energy the BH has lost, 
\begin{equation}
\Edf={\rm max}_t\,[E_{\rm orb}(0)-E_{\rm orb}(t)], 
\end{equation}
where
\begin{equation}
E_{\rm orb}=\tfrac12\mbh v_\bullet^2+\mbh\Phi_{\rm gas}(\mathbf r_\bullet)
\end{equation}
is the BH's kinetic energy in the initial rest frame of the star plus its interaction energy with all of the gas, and $\Phi_{\rm gas}$ is the softened gravitational potential of the gas at the BH position. The nuclear release $\Enuc(t)$ is the time integral of the heating rate outside the softening sphere. Its value at the end of the run is the total release. We define the energy released during the encounter, $\Eenc$, as the value of $\Enuc$ at $5\,\tdyn$, about one dynamical time after the BH has settled at the stellar center (Sec.~\ref{sec:representative}). This follows \citet{Lai1993}, who compare the energy dissipated during an encounter with the binding energy.

To quantify the bound mass and whether the star is disrupted, we compute the gas mass bound to the star alone and to the black hole star system, in units of the initial stellar mass,
\begin{equation}
f_\star=\frac{1}{\mstar}\int_{{\rm Be}_\star<0}\rho\,dV,\,\,\,
f_{\rm sys}=\frac{1}{\mstar}\int_{{\rm Be}_{\rm sys}<0}\rho\,dV,
\label{eq:fbound}
\end{equation}
where ${\rm Be}_\star=\tfrac12|\mathbf{v}-\mathbf{v}_\star|^2+h+\Phi_\star$ and ${\rm Be}_{\rm sys}=\tfrac12|\mathbf{v}-\mathbf{v}_{\rm cm}|^2+h+\Phi_\star+\Phi_\bullet$ are the Bernoulli parameters relative to the star and to the whole system. Here $h$ is the specific enthalpy, $\Phi_\star$ is the gas self-gravity potential, $\Phi_\bullet$ is the softened BH potential, $\mathbf{v}_\star$ is the bulk velocity of the stellar core, and $\mathbf{v}_{\rm cm}$ is the center-of-mass velocity of the BH and gas. Thus $f_\star$ measures the gas bound by its own gravity alone, while $f_{\rm sys}$ also includes gas bound by the BH.

\begin{table*}[!t]
\centering
\footnotesize\setlength{\tabcolsep}{6pt}
\caption{Stellar models. $f_{\rm c}$ is the mass fraction of the convective core, $\rho_{\rm c}$ and $T_{\rm c}$ are the central density and temperature, $\bbeta$ is the pressure-weighted gas fraction, $\Ebind$ is the binding energy of Eq.~(\ref{eq:virial}), and $\tdyn=(\rstar^3/G\mstar)^{1/2}$. $\bbeta$ and $\Ebind$ are measured from the initial snapshot.}
\label{tab:runs}
\begin{tabular}{lccccccccccc}
\hline\hline
model & $\mstar$ & $\rstar$ & $X$ & $Y$ & $Z$ & $f_{\rm c}$ & $\rho_{\rm c}$ & $T_{\rm c}$ & $\bbeta$ & $\Ebind$ & $\tdyn$\\
 & ($\msun$) & ($\rsun$) & & & & & (${\rm g\,cm^{-3}}$) & ($10^7$\,K) & & ($10^{50}$\,erg) & (hr)\\
\hline
Solar 100 (fiducial) & 100 & 16.0 & 0.700 & 0.280 & 0.020 & 0.80 & 1.08 & 3.7 & 0.560 & 8.97 & 2.83\\
SMC 100 & 100 & 14.0 & 0.745 & 0.253 & 0.0021 & 0.80 & 1.44 & 4.1 & 0.574 & 10.1 & 2.32\\
He-rich $Y=0.43$ & 100 & 16.0 & 0.550 & 0.430 & 0.020 & 0.80 & 1.07 & 3.8 & 0.510 & 8.08 & 2.83\\
He-rich $Y=0.78$ & 100 & 15.4 & 0.200 & 0.780 & 0.020 & 0.80 & 1.36 & 4.3 & 0.378 & 6.37 & 2.67\\
Solar 500 & 500 & 42.0 & 0.700 & 0.280 & 0.020 & 0.95 & 0.32 & 3.6 & 0.284 & 42.8 & 5.39\\
\hline
\end{tabular}
\end{table*}

\section{Results}
\label{sec:results}

\subsection{A representative collision}
\label{sec:representative}

Figure~\ref{fig:process} shows the fiducial star 
strucked by BHs with two mass ratios, $q=0.10$ ($10\,\msun$) and $q=0.20$ ($20\,\msun$). Rows~(a) and~(b) show the density for $q=0.10$ and $q=0.20$, respectively. Row~(c) shows the SMC star at the same $q=0.20$ but with only 1/10 the CNO catalyst, and row~(d) shows the nuclear heating rate around the BH for the collision in row~(b).

Each collision begins with the BH at 5 tidal radii from the stellar center, falling inward on a radial orbit. For $q=0.20$, the BH enters the star at $1.8\,\tdyn$, crosses the center at $2.3\,\tdyn$, experiences strong dynamical friction, and eventually falls back and settles at the center by $4\,\tdyn$. The orbital energy of the BH heats the gas and loosens the star. Meanwhile, the gas that gathers around the BH is compressed to nearly a hundred times the initial central density.

Nuclear burning changes the disruption picture substantially. The compressed gas, both in the shocked star (see the insets) and near the BH, burns at the hot-CNO limit. The stellar core then expands and drives the envelope outward. As the gas expands and its density drops, the burning region shrinks back to the vicinity of the BH. The white contours in rows~(a)--(c) enclose gas with a negative Bernoulli parameter in the gravity of the gas alone, i.e., gas bound to the star without the help of the BH. By the last frame of row~(b) this region has fragmented: the star is disrupted, and its self-gravity can no longer hold the gas. The $q=0.10$ collision in row~(a) follows the same sequence with less compression. Its burning rate is an order of magnitude weaker, the star survives, and the BH ends at the center of a black hole star. The SMC star of row~(c) is hit by the same $20\,\msun$ BH as in row~(b) and shows similar gas and BH dynamics during the first passage. With a tenth of the catalyst, it burns four times more slowly, and after $50\,\tdyn$ a dense remnant still surrounds the BH where the fiducial star has been disrupted.

If the enhanced nuclear reaction is neglected, the same collision 
would produce a black hole star. The BH sinks and settles at the center as before, but there is no additional pressure to lift the gas it has compressed. The mass bound by the gas alone stops falling at $10\,\tdyn$ and stays at $0.64\,\mstar$ until the end of the run at $30\,\tdyn$, compared with $0.16\,\mstar$ in the run with burning (Appendix~\ref{app:feedback}). This is the drag-only behavior that \citet{Shi2026} find up to their limit of $\mbh\simeq0.2\,\mstar$, and it identifies burning as the cause of the disruption in row~(b). Nuclear feedback lowers the BH mass above which a $100\,\msun$ star is destroyed.

\subsection{Self-regulated nuclear burning}
\label{sec:clamp}

Once the BH settles at the center, we classify the outcome by $f_\star$ (Eq. \ref{eq:fbound} in ~\S\ref{sec:setup}). A star is disrupted if $f_\star$ continues to fall long after the burning has weakened, and becomes a black hole star if $f_\star$ stabilizes. The three black hole stars stabilize at $f_\star\ge0.48$, while the four disrupted stars fall to $f_\star\le0.19$ (Table~\ref{tab:full}, Fig.~\ref{fig:histories}c). In a black hole star, $f_\star$ can be low because the BH holds much of the envelope, and $f_\star$ does not include the BH potential. The no-feedback runs at $q=0.20$ keep $f_\star=0.64$ ($100\,\msun$) and $0.38$ ($500\,\msun$), and the SMC run $0.48$, yet all three keep $f_{\rm sys}\ge0.86$ and have stopped losing mass.

In Fig.~\ref{fig:histories}, we present the detailed histories of the collisions over the first $20\,t_{\rm dyn}$. As the BH plunges into the star, a shock propagates through it and enhances the nuclear burning. In panel~(a), $L_{\rm nuc}$ rises from its unperturbed value of $\sim10^{39}\,{\rm erg\,s^{-1}}$ to a peak within a dynamical time after the first passage. The increase is $7$--$8$ orders of magnitude at $q=0.20$ and $6$--$7$ at $q=0.10$. For the disrupted runs (dashed lines), the heating rate falls by about 2 orders of magnitude from its peak to $20\,\tdyn$. In panel~(b), the cumulative release in the disrupted runs rises steeply as the BH enters and settles in the star, then plateaus at $1.0$--$1.35\,\Ebind$ (Table~\ref{tab:full}), with most of the energy released by $10\,\tdyn$. 

The plateau does not mean the fuel is exhausted. The no-feedback control of the same collision reaches one binding energy at $4.1\,\tdyn$ and continues climbing to $29.7\,\Ebind$ by $30\,\tdyn$. With nuclear feedback, the star releases about one binding energy, which is enough to expand the compressed gas out of 
the enhanced burning regime. The burning then diminishes. The yield is therefore self-regulated. In panel~(c),$f_\star$ in the disrupted runs continues to fall long after the release has flattened. In the fiducial collision it drops from $0.64$ at $5\,\tdyn$ to $0.35$ at $20\,\tdyn$, while for the $Y=0.78$ and $500\,\msun$ stars it reaches zero by $10$--$20\,\tdyn$.

In the runs that leave a black hole star (solid lines), the heating rate in panel~(a) rises to a peak as in the disrupted runs, but declines far more slowly, taking about $35\,\tdyn$ to fall by an order of magnitude. The cumulative release in panel~(b) therefore continues to rise. By $20\,\tdyn$ it has reached only $0.17$--$0.67\,\Ebind$ and is still climbing. Meanwhile,$f_\star$ in panel~(c) settles within a few dynamical times to $0.92$ and $0.84$ in the $q=0.10$ collisions, and decreases only slowly for the SMC star.

The physical origin of this trend is quantified in Table~\ref{tab:full}. The reaction rate factor $X_{\rm H}X_{\rm CNO}$ differs by a factor of 9 between the SMC and the solar-metallicity star, yet the $100\,\msun$ stars release $8.6$--$9.4\times10^{50}\,{\rm erg}$ at $q=0.20$ with a spread of only $9\%$. The SMC star has a tenth of the catalyst of the fiducial star with almost the same energy release. The yield therefore does not follow \textbf{the reaction} rate. It follows the drag energy $\Edf$, the orbital energy the BH transfers to the gas. At $q=0.20$, the nuclear release is $0.39$--$0.47\,\Edf$ for the $100\,\msun$ stars and $0.30\,\Edf$ at $500\,\msun$. Nuclear burning stops once the accumulated release reaches about half of the compression work done by the BH. At $q=0.10$ the ratio drops to $0.13$--$0.31$, because the compression is too mild to have a significant effect.

\begin{table}
\centering
\footnotesize\setlength{\tabcolsep}{4.5pt}
\caption{Detailed information of the runs. $\Edf$ is the largest orbital energy the BH loses, $\Eenc$ the nuclear release by $5\,\tdyn$, and $\Enuc$ is the total release, all in units of $\Ebind$. $f_{\rm sys}$ and $f_\star$ (Sec.~\ref{sec:setup}) are measured at the end of the run ($50\,\tdyn$). Outcomes are defined in Section~\ref{sec:clamp}.}
\label{tab:full}
\begin{tabular}{lcccccc}
\hline\hline
model & $q$ & $\Edf$ & $\Eenc$ & $\Enuc$ & $f_{\rm sys}$ & $f_\star$\\
\hline
\multicolumn{7}{c}{\textit{black hole stars}}\\
Solar 100 & 0.10 & 1.04 & 0.04 & 0.32 & 0.97 & 0.91\\
Solar 500 & 0.10 & 1.85 & 0.01 & 0.23 & 0.95 & 0.83\\
SMC 100   & 0.20 & 2.27 & 0.16 & 0.93 & 0.86 & 0.48\\
\hline
\multicolumn{7}{c}{\textit{disrupted}}\\
Solar 100        & 0.20 & 2.19 & 0.55 & 1.03 & 0.72 & 0.19\\
He-rich $Y=0.43$ & 0.20 & 2.47 & 0.68 & 1.09 & 0.54 & 0.01\\
He-rich $Y=0.78$ & 0.20 & 3.48 & 0.99 & 1.35 & 0.11 & 0.00\\
Solar 500        & 0.20 & 4.40 & 0.80 & 1.34 & 0.07 & 0.00\\
\hline
\multicolumn{7}{c}{\textit{no feedback (all black hole stars)}}\\
Solar 100 & 0.10 & 1.16 & \nodata & (1.1)  & 0.98 & 0.93\\
Solar 100 & 0.20 & 2.53 & \nodata & (29.7) & 0.91 & 0.64\\
Solar 500 & 0.10 & 1.86 & \nodata & (0.7)  & 0.95 & 0.85\\
Solar 500 & 0.20 & 4.90 & \nodata & (50.9) & 0.87 & 0.38\\
\hline
\end{tabular}
\end{table}

\subsection{The timing of the release determines the outcome}
\label{sec:criterion}

\begin{figure}
\centering
\includegraphics[width=\columnwidth]{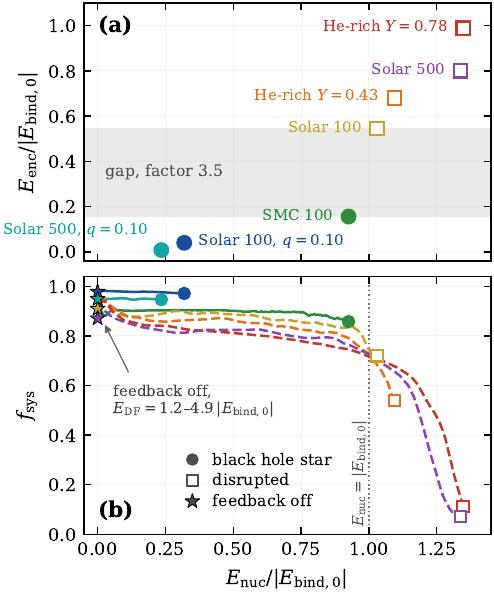}
\caption{Outcome versus total nuclear release in units of the binding energy, for every run with burning ($q=0.20$ unless marked). (a) Energy released during the encounter. The shaded band is the gap between the black hole stars and the disrupted stars. (b) Mass bound to the BH and gas together, traced along each run to the symbol at its end. Colors identify the runs as labeled in (a). Star symbols mark the no-feedback controls, which deposit no nuclear energy and take the color of their runs with burning.}
\label{fig:criterion}
\end{figure}

At $q \le 0.2$, the drag energy alone does not determine the outcome. It is $2.2$--$4.9\,\Ebind$ in the $q=0.20$ collisions, within $10$--$20\%$ of the $M_{\bullet}|\Phi(0)|$ that \citet{Shi2026} estimate for a BH sinking to the stellar center. Radiation pressure lowers $\Ebind$ by the factor $\bbeta$ (Eq.~\ref{eq:virial}). Although the drag energy already exceeds $\Ebind$ at $q=0.1$, the star can still survive. The $500\,\msun$ case even absorbs $4.9\,\Ebind$ and remains a black hole star with $f_{\rm sys}=0.87$ (Fig.~\ref{fig:criterion}b). As the BH falls from rest at infinity, energy conservation maintains the total system energy at $-\Ebind$ throughout the collision in the absence of nuclear feedback. Drag converts the BH's orbital energy into heat, but the BH's gravity deepens the potential by an equal amount as it settles. Nuclear burning is the only source of extra energy. The merged system can retain only about $\Ebind$ of it before becoming unbound, indicating that the threshold is set by the star's binding energy rather than the far larger drag energy.

Energy conservation constrains only the total energy of the system. While a purely macroscopic perspective might suggest that no star can ever be disrupted, local effects can still drive disruption. Some gas can still escape with positive energy, while the BH and the gas compressed around it become more tightly bound by the same amount. Locally, the drag heat drives the compressed gas outward just as nuclear heat does. Therefore, even without burning the collision removes a portion of the mass that grows with the BH mass (Fig.~\ref{fig:criterion}b and \citealt{Shi2026}). In this work, the no-feedback runs survive because the mass ratio is small ($q\le0.2$), whereas heavier BHs ($\mbh\gtrsim30\,\msun$) expand the envelope and drive significant mass loss even without burning.

The total nuclear energy release nevertheless does not separate the outcomes (Fig.~\ref{fig:criterion}a). The SMC star survives on $0.93\,\Ebind$ and is still burning when the run ends, while the fiducial star is destroyed on $1.03$. The distinction lies in when the energy is released within the star. Measured by $\Enuc/L_{\rm pk}$, the disrupted stars release their energy within $1.8$--$2.7\,\tdyn$ and the survivors over at least $9\,\tdyn$. Heat released within a few dynamical times drives the envelope out before the star can respond, whereas heat released over ten or more dynamical times is absorbed by a star that expands and re-virializes. \citet{Ko2022} report the same behavior for energy injected into the envelopes of massive stars, which erupt only when the injection rate exceeds the binding energy of the envelope per dynamical time. \citet{Lai1993} judge the outcome of a stellar collision by the energy dissipated during the encounter relative to the binding energy $\tfrac12\beta|W|$. We apply the same criterion to the nuclear energy released during the encounter (Sec.~\ref{sec:setup}). Normalized to the binding energy, $\Eenc$ is $0.01$--$0.16$ for the survivors and $0.55$--$0.99$ for the disrupted stars (Fig.~\ref{fig:criterion}a). A star that receives more than half its binding energy during the encounter is destroyed. One that receives less than a quarter re-virializes and survives, even if it absorbs nearly its binding energy over the following tens of dynamical times. The values of $\Eenc$ are lower bounds because burning inside the softening sphere is switched off. Halving the softening length raises $\Eenc$ from $0.55$ to $0.99\,\Ebind$ in the $q=0.20$ fiducial collision and from $0.04$ to $0.22$ at $q=0.10$, but neither collision changes its outcome (Appendix~\ref{app:convergence}).

The total release instead sets how much of the merged system stays bound. Fig.~\ref{fig:criterion}b follows $f_{\rm sys}$ along each run. The collision itself removes $2$--$5\%$ of the stellar mass at $q=0.10$ and $9$--$14\%$ at $q=0.20$. It does the same in the no-feedback controls, which absorb $1.2$--$4.9\,\Ebind$ of drag energy. The bound fraction then falls by less than $0.1$ while the release grows to $0.8\,\Ebind$, and drops steeply once the release approaches $\Ebind$. The four disrupted runs differ in mass, composition, and drag energy, yet all have $f_{\rm sys}=0.72$--$0.74$ when their release crosses $\Ebind$. The $Y=0.78$ and $500\,\msun$ stars, whose release exceeds $\Ebind$ by a third, end with $0.07$--$0.11$. The debris of the fiducial star, at $1.03\,\Ebind$, is thus marginally bound and spread over tens of stellar radii when the run ends. We do not follow its long-term fallback. The SMC star, at $0.93\,\Ebind$, stops just short of this drop and keeps $0.86$.

Structure and composition enter in two ways. Helium and mass act through the binding energy, consistent with the sensitivity of collision outcomes to stellar structure found for inert stars \citep{Freitag2005,Vynatheya2024}. Helium raises $\mu$ and lowers $\bbeta$, so by Eq.~(\ref{eq:virial}) it lowers $\Ebind$. For $Y=0.43$, $\Eenc$ is $12\%$ larger than in the fiducial star while the binding energy is $10\%$ smaller, so $\Eenc/\Ebind$ rises from $0.55$ to $0.68$. At $30\,\tdyn$ only $1\%$ of the mass remains self-bound, against $16\%$ in the fiducial star. At $Y=0.78$, $\Eenc$ reaches $0.99\,\Ebind$, the largest of any run, and nothing remains bound to the gas by $20\,\tdyn$. The $500\,\msun$ star, the softest model with $\bbeta=0.28$, absorbs $0.80\,\Ebind$ and is the most completely destroyed. The catalyst abundance enters instead through the timing. The SMC star has the same $\bbeta$ and nearly the same binding energy and total release as the fiducial star, but it releases only one-sixth of its total during the encounter, whereas the fiducial star releases half.

\subsection{The black hole stars}
\label{sec:bhstars}

All three black hole stars remain bound, with $f_{\rm sys}=0.86$--$0.97$ and the BH settled near the center of the self-bound gas. The SMC star is the least bound, with $f_\star=0.48$, but the BH still holds most of its envelope. They are not quiescent. The BH holds the surrounding gas at $50\,{\rm g\,cm^{-3}}$ and $1.5\times10^8\,{\rm K}$, where the CNO cycle burns close to its cap, 8 orders of magnitude faster than at the original stellar center. Their burning declines roughly as $t^{-1}$. Between $10$ and $50\,\tdyn$, it runs at $2\times10^{44}$--$5\times10^{45}\,{\rm erg\,s^{-1}}$, four orders of magnitude above the Eddington luminosity of the star. By the end of the runs the $q=0.10$ stars have released $0.23$--$0.32\,\Ebind$ and the SMC star $0.93$, and all are still rising. Burning regulates itself here as well. Without feedback, the $q=0.10$ collisions release three to four times more by $30\,\tdyn$ while keeping nearly the same bound mass (Appendix~\ref{app:feedback}). At $q=0.10$, the hot gas lies partly inside the softening sphere, where burning is switched off, so these rates are lower limits.

\section{Summary and discussion}
\label{sec:discussion}

When a BH with $q\le0.2$ collides with a star, dynamical friction alone does not disrupt it. The BH dissipates $1.0$--$4.9\,\Ebind$ into the envelope, but its gravity deepens the potential by a comparable amount. Enhanced nuclear burning adds a new source
of auxiliary power. Once the cumulative release approaches the binding energy, the merged system comes apart, whatever the mass, composition, or drag energy of the collision. Gas compressed around the BH burns at the $\beta$-limited hot-CNO rate and self-regulates near one binding energy, so the timing of the release determines the outcome. A star receiving more than half its binding energy within a few dynamical times is disrupted, and one receiving less than a quarter survives as a black hole star. This feedback lowers the retention threshold below the drag-only value of $q\simeq0.2$ \citep{Shi2026}, most strongly in weakly bound stars, rich in helium or very massive (Eq.~\ref{eq:virial}). Metal-poor stars are the most resilient, and Population~III stars ($X_{\rm CNO}\lesssim10^{-4}$) collide as if they were inert. Our models span a wide range of stellar compositions and masses to systematically test this criterion. 

The retention threshold also caps the BHs that can merge inside stars. A black hole star that captures a second BH drags the inner binary toward a gravitational-wave merger, and the remnant can seed further generations \citep{Hu2026,Ma2026,Cantiello2026}. Each captured secondary must satisfy $\mbh\lesssim q_{\rm c}\mstar$, with $q_{\rm c}\simeq0.1$--$0.2$, so the secondary mass is capped even as the primary grows through successive collisions. Secondary masses observed by LIGO--Virgo--KAGRA avoid $\approx44$--$116\,\msun$ while some primaries lie inside this range \citep{2026Natur.652..874T}. This gap is usually attributed to pair-instability supernovae, but its lower edge could also reflect the capture limit in hosts of $220$--$440\,\msun$ formed in AGN disks \citep{Cantiello2021,AliDib2023} or through runaway collisions in dense clusters \citep{Rantala2026}. Because burning lowers $q_{\rm c}$ in metal- and helium-rich hosts, this edge should shift with host composition, a testable signature of mergers inside stars.

Our models have several limitations. The most important is that we omit accretion onto the BH and its radiative feedback near the Eddington limit, which in reality set how close to the BH the gas can be compressed. The softening length stands in for this inner boundary, so halving it changes the energy released, but not which stars survive (Appendix~\ref{app:convergence}). Oblique encounters, radiation transport, fallback of the debris, which may power a long-lived accretion transient \citep{Cantiello2026}, and embedded binaries remain to be explored.

\begin{acknowledgments}
H.-X.J. thanks Zhenghao Xu for the useful discussion and Jing-Ze Xia for the support of GPU time. The authors thank Yanlong Shi, Qingru Hu, and Sizheng Ma for sharing their results in advance of publication.
\end{acknowledgments}

\software{\athenak\ \citep{Stone2026, Jiang2026}, \texttt{numpy}, \texttt{scipy}, \texttt{matplotlib}, \texttt{numba}}

\appendix

\section{Diagnostics, momentum and tests}
\label{app:convergence}
\begin{figure*}
\centering
\includegraphics[width=\textwidth]{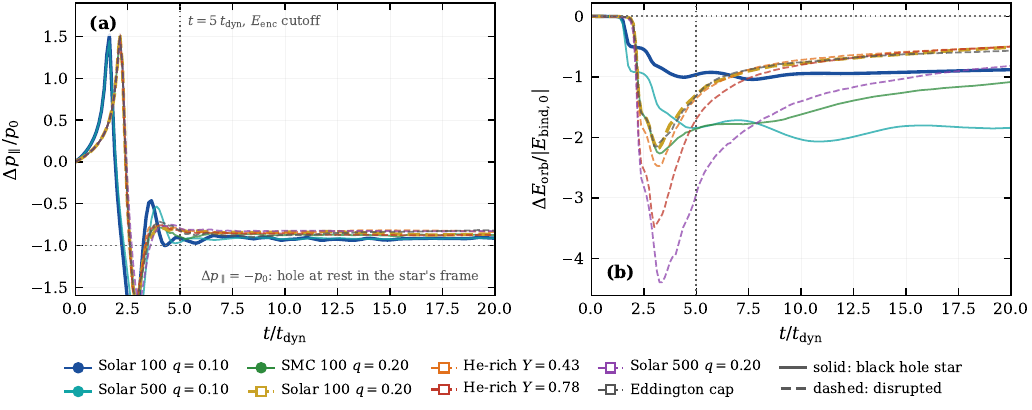}
\caption{(a) Momentum of the BH along its incoming direction and (b) its orbital energy, in the initial rest frame of the star. ``Eddington cap'' is the test run in which gas inside the softening sphere burns at up to the Eddington luminosity of the BH.\label{fig:momentum}}
\end{figure*}

The measured $\bbeta$ of the initial models agrees with the Eddington quartic $(1-\beta)/\beta^4=(\mu^2\mstar/18.1\,\msun)^2$ to better than $1\%$. The nuclear luminosity of the initial fiducial model is $1.4\times10^{39}\,{\rm erg\,s^{-1}}$, below the Eddington-model value $4\pi cG\mstar(1-\beta)/\kappa=6.5\times10^{39}\,{\rm erg\,s^{-1}}$ that sets its radius. The central temperature of the model is therefore $10\%$ below the value that would carry that luminosity. It does not matter, because the runs last $50\,\tdyn\ll t_{\rm KH}$ and the burning rate during the encounter is set by the cap. Above $10^8$\,K the tabulated equation of state hands over to the Helmholtz equation of state of \citet{Timmes2000}, and the burning gas is a fully ionized non-degenerate gas plus radiation, where the two agree. 

Fig.~\ref{fig:momentum} resolves the momentum of the BH along its incoming direction in the initial rest frame of the star. The BH gains $1.5\,p_0$ falling to the center and loses it again within a dynamical time. It ends at $\Delta p_\parallel/p_0=-0.76$ to $-0.90$, close to the value $-1/(1+q)$ of a BH that moves with the center of mass, and the transverse kick is below $0.03\,p_0$ at the end of these head-on runs. In the black hole stars the loss of orbital energy is permanent, $\Delta E_{\rm orb}/\Ebind=-0.5$ to $-1.7$ with burning and $-5.0$ in the $500\,\msun$ control, whereas in the disrupted runs it unwinds to $-0.3$ to $-0.6$ once the envelope disperses and the interaction energy that stored it is gone.

\begin{figure*}
\centering
\includegraphics[width=\textwidth]{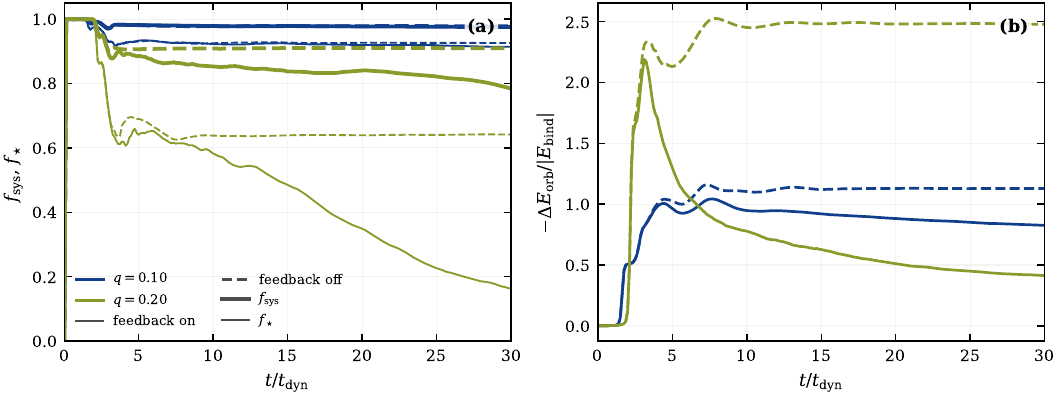}
\caption{The $100\,\msun$ collisions with nuclear feedback on (solid) and off (dashed). (a) $f_{\rm sys}$ (thick) and $f_\star$ (thin). (b) Orbital energy the BH has lost.}
\label{fig:feedback}
\end{figure*}

Fig.~\ref{fig:feedback} compares the $100\,\msun$ collisions with the nuclear feedback on and off\label{app:feedback}. In the controls the network runs as a tracer and the gas is not heated. The runs of each pair are identical through the first passage and separate once the burning peaks. With the feedback off the BH keeps the orbital energy it has given up and the bound mass settles within a few dynamical times. At $q=0.20$, the control stays flat from $10\,\tdyn$ to the end of the run, with $0.64\,\mstar$ bound by the gas alone, $0.91\,\mstar$ bound to the BH and gas together, and the BH $0.33\,\rsun$ from the center. Its twin with feedback falls to $0.16\,\mstar$ over the same interval. The same collision therefore leaves a black hole star with the burning off and destroys the star with it on. At $q=0.10$ feedback barely matters. The fiducial star keeps $0.91\,\mstar$ bound by the gas alone with burning on against $0.93$ with it off, and the $500\,\msun$ star $0.83$ against $0.85$. At $q=0.20$ the $500\,\msun$ star keeps $0.00$ against $0.38$. The heavier BH compresses the gas harder, so the loosening by burning appears at $q=0.20$ but not at $q=0.10$. Drag alone also leaves a black hole star at $500\,\msun$. The control absorbs $4.9\,\Ebind$ and keeps $87\%$ of the stellar mass bound, but only $38\%$ by the gas alone, so this star may lie close to the drag-only threshold.

The classification does not depend on the cutoff. By $4\,\tdyn$ the SMC and fiducial stars have received $0.09$ and $0.36\,\Ebind$, and by $10\,\tdyn$ $0.41$ and $0.77$. The burning inside the softening sphere is switched off, and the release it would have produced is measured from the snapshots by applying the capped rate to the gas inside one softening length. At the peak of the fiducial collision that gas holds $0.06\,\msun$ at $90\,{\rm g\,cm^{-3}}$ and $2.2\times10^8$\,K and would burn at $1.4\times10^{46}\,{\rm erg\,s^{-1}}$, $0.4$ of the heating rate. By $5\,\tdyn$ it would add $0.33\,\Ebind$ to the fiducial star, $0.97$ to the $500\,\msun$ star, $0.04$ to the SMC star and $0.13$ to the $q=0.10$ star. The values of $\Eenc$ are therefore lower bounds, for a non-accreting BH. With this addition the survivors stay below $0.19\,\Ebind$ while the disrupted stars rise above $0.86$. Halving the softening length raises $\Eenc$ by more than this estimate (see below).

\begin{figure*}
\centering
\includegraphics[width=\textwidth]{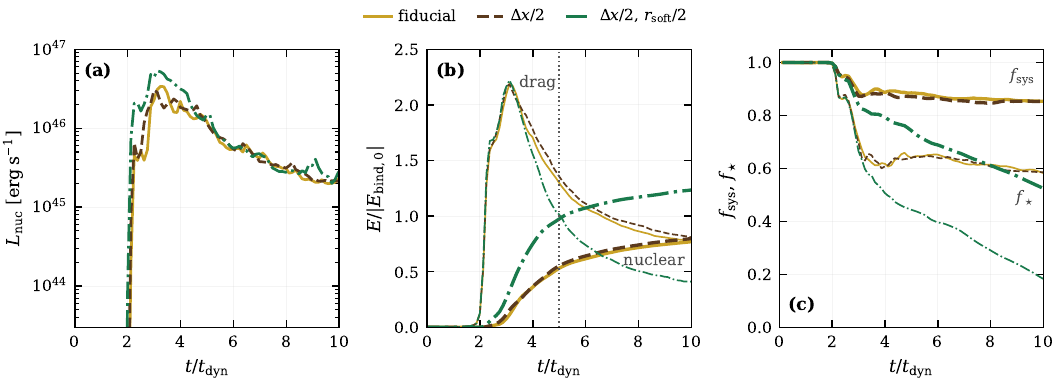}
\caption{The fiducial $q=0.20$ collision (solid), the same collision with twice the cells across the same softening length (dashed), and with twice the cells and half the softening length (dash-dotted). (a) Nuclear heating rate. (b) Nuclear release (thick) and orbital energy the BH has given up (thin). (c) $f_{\rm sys}$ (thick) and $f_\star$ (thin).\label{fig:resolution}}
\end{figure*}

Three runs test the treatment of the gas around the BH. One lets the gas inside the softening sphere burn at up to the Eddington luminosity of the BH, which is negligible beside the burning outside it. It reproduces the fiducial $\Eenc$ to $6\%$ ($\Eenc/\Ebind=0.58$ against $0.55$). At $20\,\tdyn$, where it ends, the release agrees to $2\%$, and the bound fractions are $f_{\rm sys}=0.82$ and $f_\star=0.43$, against $0.84$ and $0.35$ in the fiducial run. The two histories coincide to $15\,\tdyn$ (Fig.~\ref{fig:histories}), and the difference afterwards measures the scatter between realizations of the same collision. The second resolves the same softening length with eight cells instead of four. At this resolution the compressed gas at the stellar surface reaches a higher peak wave speed, and the run only survives the peak of the burning if the Courant number is halved, from $0.5$ to $0.25$. With that change it runs cleanly through $10\,\tdyn$. Through the first passage, to $2.5\,\tdyn$, it reproduces the trajectory of the BH, the drag energy and the bound mass of the fiducial run to $1\%$, as before. Over the full $10\,\tdyn$ (Fig.~\ref{fig:resolution}) the peak drag energy agrees to $0.3\%$, $\Eenc$ to $4\%$, with $\Eenc/\Ebind=0.57$ against $0.55$, and the release by $10\,\tdyn$ also to $4\%$, and both bound masses stay within $0.03\,\mstar$ of the fiducial run. After the close passage, the BH position departs from that in the fiducial run, reaching a separation of $0.19\,\rstar$ by $10\,\tdyn$. A close encounter is expected to amplify small differences between two numerical solutions in this way, and the difference does not carry into the bound mass or the released energy. At fixed softening length, resolving it with twice the cells does not change the outcome. The third also halves the softening length. It leaves the drag energy unchanged to $2\%$ but raises $\Eenc$ from $0.55$ to $0.99\,\Ebind$. The burning gas becomes three times denser and $1.6$ times hotter, and at $10\,\tdyn$ the star keeps $18\%$ of its mass bound by the gas alone, against $58\%$. The gas around the BH is dominated by radiation pressure, and the mass hot enough to burn grows as the inner edge of this atmosphere moves inward, so a smaller softening length releases more energy, faster. This atmosphere has no length scale of its own, so its inner edge is set here by the softening length and in reality by accretion (Sec.~\ref{sec:discussion}). It still keeps $0.73$ of the stellar mass bound when its release crosses $\Ebind$, as the four disrupted runs do, so the energy a collision releases depends on the softening length but the response of the star to it does not. We repeat the third test at $q=0.10$. It raises $\Eenc$ from $0.04$ to $0.22\,\Ebind$, and by $9.7\,\tdyn$, where it ends, the release from $0.11$ to $0.43\,\Ebind$ and the heating rate fourfold, yet both bound fractions stay within $0.01$ of the fiducial run, at $0.97$ and $0.92$, and the collision still leaves a black hole star.

\bibliography{refs}
\bibliographystyle{aasjournalv7}

\end{document}